\documentclass[12pt]{article}
\usepackage{amssymb}
\newcommand{\dmat}[3]{%
    \begin{array}{ccc}
    	#1 &  &   \\
    	 & #2 &   \\
    	 &  & #3
    \end{array}}
\newcommand{\umat}[3]{%
    \begin{array}{ccc}
    	 & #1 & #2   \\
    	 &  & #3  \\
    	 &  & 
    \end{array}}
\newcommand{\lmat}[3]{%
    \begin{array}{ccc}
    	 &  &   \\
    	#1 &  &   \\
    	#2 & #3 & 
    \end{array}}
\newcommand{\be}{\begin{eqnarray}}
\newcommand{\ee}{\end{eqnarray}}
\newcommand{\non}{\nonumber}
\newcommand{\tr}{\mathop{\rm tr}\nolimits}
\newcommand{\id}{\mathbb{I}}

\begin{document}

\begin{titlepage}
\strut\hfill UMTG--220
\vspace{.5in}
\begin{center}

\LARGE Nonstandard coproducts and the Izergin-Korepin open spin chain \\[1.0in]
\large Rafael I. Nepomechie\\[0.8in]
\large Physics Department, P.O. Box 248046, University of Miami\\[0.2in]  
\large Coral Gables, FL 33124 USA\\

\end{center}

\vspace{.5in}

\begin{abstract}
Corresponding to the Izergin-Korepin ($A_{2}^{(2)}$) $R$ matrix, there
are three diagonal solutions (``$K$ matrices'') of the boundary
Yang-Baxter equation.  Using these $R$ and $K$ matrices, one can
construct transfer matrices for open integrable quantum spin chains. 
The transfer matrix corresponding to the identity matrix $K=\id$ is
known to have $U_{q}(o(3))$ symmetry.  We argue here that the transfer
matrices corresponding to the other two $K$ matrices also have
$U_{q}(o(3))$ symmetry, but with a nonstandard coproduct.  We briefly
explore some of the consequences of this symmetry.
\end{abstract}

\end{titlepage}

\section{Introduction and summary}

The notion of coproduct is of fundamental importance in the theory of 
representations of algebras.  Given a representation of an algebra on 
a vector space $V$, the coproduct $\Delta$ determines the 
representation on the tensor product space $V \otimes V$.  For a 
classical Lie algebra, the coproduct is trivial: for any generator 
$x$, the coproduct is $\Delta(x) = x \otimes \id  + \id \otimes x$,
where $\id$ is the identity matrix. For 
quantum algebras, the situation is more interesting.  Indeed, 
consider the case $U_{q}(su(2))$, with a set of three generators 
$\{ j_{\pm} \,, h \}$ obeying
\be
\left[ h \,, j_{\pm} \right] = \pm j_{\pm} \,.
\ee
As is well known, the ``standard''  coproduct
\be
\Delta(h) &=& h \otimes \id + \id \otimes h \,, \non \\
\Delta(j_{\pm}) &=& j_{\pm}\otimes q^{h} + q^{-h} \otimes j_{\pm} \,, 
\label{standard}
\ee
is compatible with the commutation relation
\be
\left[ j_{+} \,, j_{-} \right] = {q^{2 h}-q^{-2 h}\over q - q^{-1}} \,.
\ee 
Perhaps less well-known is the fact that there is also a 
``nonstandard'' coproduct
\be
\Delta(h) &=& h \otimes \id + \id \otimes h \,, \non \\
\Delta(j_{\pm}) &=& j_{\pm}\otimes \id  + q^{h} \otimes j_{\pm} \,, 
\label{nonstandard}
\ee
which is compatible instead with the $q$-commutation relation
\be
j_{+}  j_{-} - q^{-1}  j_{-} j_{+} = {\id - q^{2h}\over 1 - q^{2}} \,.
\ee

Remarkably, both of these types of coproducts can be realized in the 
open integrable quantum spin chain constructed with the $A_{2}^{(2)}$ 
$R$ matrix \cite{IK} by choosing appropriate boundary conditions.  Let 
us briefly recall the history of this model.  Sklyanin \cite{sklyanin} 
pioneered the generalization of the Quantum Inverse Scattering Method 
\cite{QISM} to systems with boundaries, and showed that integrable 
boundary conditions can be obtained from solutions $K(u)$ of the 
boundary Yang-Baxter equation \cite{cherednik}, \cite{GZ}. This 
approach was then generalized \cite{MN1} to spin chains associated 
with general affine Lie algebras \cite{bazhanov}, \cite{jimbo}. In 
particular, for the $A_{2}^{(2)}$ case, it was found \cite{MN2} 
that there are only three diagonal solutions of the boundary 
Yang-Baxter equation:
\be
K^{(0)}(u) &=& \id = diag \Bigl( 1 \,, 1 \,, 1 \Bigr) \,, \non  \\
K^{(1)}(u) &=& diag \Bigl(  e^{-u} \,, 
{\sinh({1\over 2}(3 \eta - {i \pi\over 2} + u))\over  
 \sinh({1\over 2}(3 \eta - {i \pi\over 2} - u))} \,, e^{u} \Bigr)  
\,, \non  \\
K^{(2)}(u) &=& diag \Bigl(   e^{-u} \,, 
{\cosh({1\over 2}(3 \eta - {i \pi\over 2} + u))\over  
 \cosh({1\over 2}(3 \eta - {i \pi\over 2} - u))} \,, e^{u} \Bigr) \,,
\label{kmatrices}
\ee
where $u$ is the spectral parameter, and $\eta$ is the anisotropy 
parameter.  Let us denote the corresponding transfer matrices for open 
quantum spin chains with $N$ sites by $t^{(i)}(u)$, $i = 0 \,, 1 \,, 
2$.  (The construction of these transfer matrices is described below 
in Section 2.) It was shown in \cite{MN3}, \cite{MN4} that the 
transfer matrix $t^{(0)}(u)$ constructed with the identity matrix 
$K^{(0)}$ has $U_{q}(o(3))$ symmetry:
\be
\left[ t^{(0)}(u) \,, S^{\pm} \right] = 0 \,, \qquad 
\left[ t^{(0)}(u) \,, S^{3} \right] = 0 \,,
\ee
where the generators obey
\be
\left[ S^{3} \,, S^{\pm} \right] = \pm S^{\pm} \,, \qquad
\left[ S^{+} \,, S^{-} \right] = {q^{2 S^{3}}-q^{-2 S^{3}}\over q - q^{-1}} \,,
\ee 
and
\be
S^{\pm}= \sum_{k=1}^N q^{s^{3}_N + \cdots + s^{3}_{k+1}}\ s^{\pm}_{k} \ 
q^{-(s^{3}_{k-1} + \cdots + s^{3}_1)}\,, 
\qquad S^{3} = \sum_{k=1}^N s^{3}_k \,,
\ee
where
\be
s^{+} = {\sqrt{2\cosh \eta}} \left( \begin{array}{ccc}
                            0   & 1   & 0   \\
                            0   & 0   & 1   \\                 
                            0   & 0   & 0          
					           \end{array} \right) \,, \quad 
s^{-} = {\sqrt{2\cosh \eta}} \left( \begin{array}{ccc}
                            0   & 0   & 0    \\
                            1   & 0   & 0    \\                 
                            0   & 1   & 0    \\        
					           \end{array} \right) \,, \quad 
s^{3} =   \left( \begin{array}{ccc}
                            1      \\
                            &   0   \\                 
                            &   &   -1        
					           \end{array} \right)  \,,
\ee 
and $q = e^{\eta}$.  That is, the transfer matrix has quantum algebra
symmetry with the ``standard'' coproduct (\ref{standard}).  This is a
generalization of the observation \cite{PS}, \cite{KS} of
$U_{q}(su(2))$ symmetry for the $A_{1}^{(1)}$ case.  Batchelor and
Yung \cite{BY} later showed that the open $A_{2}^{(2)}$ spin chain can
be mapped to the problem of polymers at surfaces, and that the above
three solutions $K^{(i)}(u)$ correspond to three distinct surface critical 
behaviors.

There has remained the question: what symmetry -- if any -- do the 
transfer matrices constructed with $K^{(1)}$ and $K^{(2)}$ have? 
Naively, one expects that since $K \ne \id$, there is less 
symmetry. \footnote{This expectation holds true for the $A_{n}^{(1)}$
case \cite{DN}.  Indeed, there the diagonal $K$ matrices contain an 
additional continuous parameter $\xi$; and $K = \id$ is a 
point ($\xi \rightarrow \infty )$ of enhanced symmetry.}
However, this is {\it not} the case. We 
argue here that the transfer matrices $t^{(1)}(u)$ and $t^{(2)}(u)$ 
also have $U_{q}(o(3))$ symmetry, but with a ``nonstandard'' 
coproduct (\ref{nonstandard}):
\be
\left[ t^{(i)}(u) \,, S^{\pm} \right] = 0 \,, \qquad 
\left[ t^{(i)}(u) \,, S^{3} \right] = 0 \,, \qquad i = 1 \,, 2 \,,  
\label{mainsymmetry1}
\ee
where the generators obey
\be
\left[ S^{3} \,, S^{\pm} \right] = \pm 2 S^{\pm} \,, \qquad
S^{+} S^{-}  - q^{-2} S^{-}  S^{+} = {\id - q^{2S^{3}}\over 1 - q^{2}} \,,
\label{mainsymmetry2}
\ee 
and
\be
S^{\pm}= \sum_{k=1}^N s^{\pm}_{k} \ q^{s^{3}_{k-1} + \cdots + s^{3}_1} \,,  \qquad 
S^{3} = \sum_{k=1}^N s^{3}_k  \,,
\label{mainsymmetry3}
\ee
where
\be
s^{+} =  \left( \begin{array}{ccc}
                            0   & 0   & 1   \\
                            0   & 0   & 0   \\                 
                            0   & 0   & 0          
					           \end{array} \right) \,, \quad 
s^{-} =  \left( \begin{array}{ccc}
                            0   & 0   & 0    \\
                            0   & 0   & 0    \\                 
                            1   & 0   & 0    \\        
					           \end{array} \right) \,, \quad 
s^{3} =   \left( \begin{array}{ccc}
                            1      \\
                            &   0   \\                 
                            &   &   -1        
					           \end{array} \right)  \,,
\label{mainsymmetry4}				   
\ee
and $q = e^{4\eta}$.  Knowledge of such symmetry is essential for 
understanding important features of the models such as degeneracies of 
the spectrum and the Bethe Ansatz solution.  Eqs.  
(\ref{mainsymmetry1}) - (\ref{mainsymmetry4}) are the main results of 
this Letter.  In Section 2 we provide some pertinent details about the 
construction and symmetry of the models, and we conclude in Section 3 
with a brief discussion.

\section{Some details}

In this Section, we briefly review the construction of the transfer 
matrices, and outline the argument for their symmetry.  The solution 
$R(u)$ of the Yang-Baxter equation found by Izergin and Korepin 
\cite{IK}, which corresponds \cite{bazhanov},\cite{jimbo} to the case 
$A_{2}^{(2)}$, can be written in the following form \cite{KS0}, 
\cite{MN4}
\be
R(u) = \left(
    \begin{array}{c|c|c}
    	\dmat{c}{b}{d} & \lmat{e}{\relax}{g} & \lmat{\relax}{f}{\relax}  \\
		\hline
		\umat{\bar e}{\relax}{\bar g} & \dmat{b}{a}{b} & \lmat{g}{\relax}{e}  \\
		\hline
    	\umat{\relax}{\bar f}{\relax} & \umat{\bar g}{\relax}{\bar e} & 
    	\dmat{d}{b}{c}
    \end{array}
\right)
\ee
where
\be 
a &=& \sinh (u - 3\eta) - \sinh 5\eta + \sinh 3\eta + \sinh \eta \,, \quad\quad
b =  \sinh (u - 3\eta) + \sinh 3\eta  \,,  \non \\
c &=& \sinh (u - 5\eta) + \sinh \eta  \,, \quad\quad\quad\quad\quad\quad\quad\quad
d =  \sinh (u - \eta) + \sinh \eta \,,  \non \\
e &=& -2 e^{-{u\over 2}} \sinh 2\eta \ \cosh ({u\over 2} - 3\eta)  \,, 
\quad\quad\quad\quad\quad
\bar e = -2 e^{{u\over 2}} \sinh 2\eta \ \cosh ({u\over 2} - 3\eta)  \,, \non \\
f &=& -2 e^{-u + 2\eta} \sinh \eta \ \sinh 2\eta -  e^{-\eta} \sinh 4\eta \,, 
\quad
\bar f = 2 e^{u - 2\eta} \sinh \eta \ \sinh 2\eta -  e^{\eta} \sinh 4\eta \,,
\non \\
g &=& 2 e^{-{u\over 2} + 2\eta} \sinh {u\over 2} \ \sinh 2\eta \,, 
\quad\quad\quad\quad\quad\quad\quad
\bar g = - 2 e^{{u\over 2} - 2\eta} \sinh {u\over 2} \ \sinh 2\eta \non \,.
\ee 
It has the regularity property $R(0) \propto {\cal P}$, where ${\cal P}$ is the 
permutation matrix, as well as unitarity, $PT$ symmetry, and crossing 
symmetry
\be
R_{12}(u) = V_1 \ R_{12}(-u - \rho)^{t_2}\  V_1 
=  V_2^{t_2} \ R_{12}(-u - \rho)^{t_1}\  V_2^{t_2}  \,, 
\ee
where the crossing matrix $V$ is given by
\be 
V= \left( \begin{array}{ccc}
                      &   &  -e^{-\eta}   \\
                      &   1               \\                 
                      -e^{\eta}           
					  \end{array} \right) \,,
\ee
and $\rho = -6\eta - i \pi$.

Given a solution $K(u)$ of the boundary Yang-Baxter equation, a 
corresponding transfer matrix $t(u)$ for an open integrable quantum 
spin chain with $N$ sites is given by \cite{sklyanin}, \cite{MN1}, 
\cite{DN}
\be
t(u) = \tr_{0} M_{0} K_{0}(-u-\rho)^{t_{0}} {\cal T}_{0}(u) \,,
\ee
where
\be
{\cal T}_{0}(u) = T_{0}(u)\ K_{0}(u)\ \hat T_{0}(u) \,,
\label{calT}
\ee
with
\be
T_{0}(u) = R_{0N}(u) \cdots R_{01}(u) \,, \qquad 
\hat T_{0}(u) = R_{10}(u) \cdots R_{N0}(u) \,,
\ee
and 
\be
M = V^{t} V = diag \Bigl(  e^{2 \eta} \,, 1 \,, e^{-2 \eta}  \Bigr) \,. 
\ee 
Indeed, the transfer matrix forms a one-parameter commutative family
$\left[ t(u) \,, t(v) \right] = 0$, which contains the Hamiltonian 
${\cal H}$,
\be
{\cal H} \propto {d\over du}t(u) \Big\vert_{u=0} \,.
\label{hamiltonian}
\ee

For the three $K$ matrices $K^{(i)}(u)$ given in Eq.  
(\ref{kmatrices}), we denote by $t^{(i)}(u)$ the corresponding 
transfer matrices, and by ${\cal H}^{(i)}$ the corresponding 
Hamiltonians.  We now restrict our attention to the cases $i=1 \,, 2$.  
For 2 sites ($N=2$), we have checked the $U_{q}(o(3))$ symmetry 
(\ref{mainsymmetry1}) - (\ref{mainsymmetry4}) of the transfer matrix 
by direct computation.  In particular, Eq.  (\ref{hamiltonian}) 
implies that the 2-site Hamiltonian also has this symmetry.  For 
general $N$, the Hamiltonian is given by a sum of 2-site Hamiltonians 
plus boundary terms.  It follows that, for general $N$, the 
Hamiltonian ${\cal H}^{(i)}$ has $U_{q}(o(3))$ symmetry
\be
\left[ {\cal H}^{(i)} \,, S^{\pm} \right] = 0 \,, \qquad 
\left[ {\cal H}^{(i)} \,, S^{3} \right] = 0  \,, \qquad i = 1 \,, 2 \,,  
\ee
where the symmetry generators obey 
(\ref{mainsymmetry2}) - (\ref{mainsymmetry4}).  We have also checked the 
symmetry (\ref{mainsymmetry1}) of the transfer matrix for 3 sites 
($N=3)$ by direct computation, and we conjecture that it holds for 
general $N$.

We remark that the symmetry generators $S^{\pm}$, $S^{3}$ 
defined in (\ref{mainsymmetry3}), (\ref{mainsymmetry4}) lie in the fundamental 
algebraic structures of QISM. Indeed, note the asymptotic behavior of 
the $R$ and $K$ matrices for $u \rightarrow \infty$ :
\be
R(u) & \sim & e^{u} R^{+} + R^{++} + O(e^{-u}) \,, \\
K^{(i)}(u) & \sim & e^{u} K^{(i)+} + K^{(i)++} + O(e^{-u}) 
\,, \qquad i = 1 \,, 2 \,,  
\ee
where $R^{+} \,, R^{++} \,, K^{(i)+} \,, K^{(i)++}$ are independent of $u$. 
It follows that the quantity ${\cal T}^{(i)}(u)$ defined as in Eq. (\ref{calT}) 
has the asymptotic behavior for $u \rightarrow \infty$
\be
{\cal T}^{(i)}(u) \sim e^{(2N + 1)u} {\cal T}^{(i)+} + e^{2 N u} {\cal 
T}^{(i)++} + \ldots \,,
\ee 
where ${\cal T}^{(i)+} \,, {\cal T}^{(i)++}$ are independent of $u$. 
The basic observation is that the generators $S^{\pm}$ lie in the 
antidiagonal corners of ${\cal T}^{(i)++}$ (viewed as a $3 \times 3$ 
auxiliary-space matrix, with operator-valued entries):
\be
{\cal T}^{(i)++} = \left( \begin{array}{ccc}
    0       & 0       & S^{-}  \\ \relax
    0       & *       & *  \\
	S^{+}   & *       & *  
\end{array} \right) \,.
\ee 
We expect that this observation will be useful for formulating a QISM 
proof of the symmetry (\ref{mainsymmetry1}). 

\section{Discussion}

One immediate consequence of the symmetry which we have uncovered is 
the explanation of degeneracies in the spectrum for finite $N$. For 
instance, consider the pseudovacuum vector $\omega = \left( 
\begin{array}{c}
	1 \\
	0 \\
	0
	\end{array} \right)^{\otimes N}$,
\be
t^{(i)}(u)\ \omega = \Lambda^{(i)}(u)\ \omega \,, \qquad i = 1 \,, 2 \,, 
\ee
where $\Lambda^{(i)}(u)$ is the corresponding pseudovacuum eigenvalue.  
Commutativity of the transfer matrix with $S^{-}$ implies that the 
vectors $(S^{-})^{n} \omega $ for $n = 1\,, 2 \,, \ldots \,, N$ are 
also eigenvectors of the transfer matrix with the same eigenvalue.
Moreover, we observe that each site carries a {\em reducible} 
representation of the $U_{q}(o(3))$ algebra, namely 
${\bf 2} \oplus {\bf 1}$ (instead of ${\bf 3})$, implying the degeneracy pattern
$({\bf 2} \oplus {\bf 1})^{\otimes N}$.

Note that the pseudovacuum vector $\omega$ is annihilated by $S^{+}$;
that is, $S^{+} \omega = 0$.  We expect that all Bethe Ansatz states
(which can presumably be constructed by applying appropriate
creation-like operators to $\omega$) are such highest-weight states. 
(See, e.g., \cite{FT}, \cite{MN3}, \cite{DVGR}, \cite{karowski}.)

Finally, we remark that we have considered here only the first 
of the infinite family of models $A_{2n}^{(2)}$, $n = 1 \,, 2 \,, \ldots$.
For these $R$ matrices \cite{bazhanov}, \cite{jimbo}, there are again 
only three distinct diagonal solutions of the boundary Yang-Baxter 
equation: $K^{(0)}= \id$ \cite{MN2}, and $K^{(1)} \,, K^{(2)}$  given 
in \cite{BFKZ}. The transfer matrix constructed with $K^{(0)}$ has 
\cite{MN3} the symmetry $U_{q}(o(2n+1))$ with the standard coproduct. We 
expect that the transfer matrices constructed with
$K^{(1)}$ and $K^{(2)}$ also have $U_{q}(o(2n+1))$ symmetry, but 
with a nonstandard coproduct. We hope to report on this and related 
matters in a future publication.

\section*{Acknowledgments}

I am grateful to O. Alvarez for his helpful comments.
This work was supported in part by the National Science Foundation 
under Grant PHY-9870101.

\end{document}